\documentclass[prl,twocolumn,showpacs,floatfix]{revtex4}
\usepackage{graphics}
\usepackage{graphicx}
\begin{document}
\newcommand{\RR}{\mathrm{\mathbf{R}}}
\newcommand{\rr}{\mathrm{\mathbf{r}}}
\newcommand{\defin}{\stackrel{def}{=}}

\title{Re-entrant ferromagnetism in a generic class of diluted magnetic semiconductors}
\author{M.J. Calder\'on and S. {Das Sarma}}

\affiliation{Condensed Matter Theory Center, Department of Physics,
University of Maryland, College Park, MD 20742-4111}

\date{\today}


\begin{abstract}

Considering a general situation where a semiconductor is doped by magnetic impurities leading to a carrier-induced ferromagnetic exchange coupling between the impurity moments, we show theoretically the possible generic existence of three ferromagnetic transition temperatures, $T_1 > T_2 > T_3$, with two distinct ferromagnetic regimes existing for  $T_1 > T > T_2$ and $T < T_3$.  Such an intriguing re-entrant ferromagnetism, with a paramagnetic phase ($T_2 > T > T_3$)  between two ferromagnetic phases, arises from a subtle competition between indirect exchange induced by thermally activated carriers in an otherwise empty conduction band versus the exchange coupling existing in the impurity band due to the bound carriers themselves.  We comment on the possibility of observing such a re-entrance phenomenon in diluted magnetic semiconductors and magnetic oxides.

\end{abstract}

\maketitle
\pagebreak
The large class of materials comprised of diluted magnetic
semiconductors and magnetic oxides (DMS) has been widely studied in
the recent years for their potential in spintronic applications as well as for the fundamental physics of carrier-mediated ferromagnetism in semiconductors. They
have ferromagnetic (FM) critical temperatures $T_{\rm C}$ ranging from below (e.g. (Ga,Mn)As~\cite{ohno96}) to above room temperature (e.g.  doped magnetic oxides as TiO$_2$~\cite{matsumoto01}), and some others never show long range FM order (e.g. most Mn-doped II-VI semiconductor alloys are spin glasses) \cite{furdyna88}. FM is usually ascribed to carrier-mediated mechanisms and depends on many different parameters (e.g. carrier density $n_c$, magnetic impurity density $n_i$, magnetic coupling between the ion moment and the electron (or hole) spins $J$, and details of disorder) which vary greatly from system to system and sometimes from sample to sample. When the concentration of magnetic impurities is larger than a few percent, direct exchange (i.e. not carrier-mediated) can also be significant. This exchange is antiferromagnetic in many cases, like the Mn-doped III-V~\cite{vonmolnar91} and II-VI semiconductors~\cite{larson88}, and ferromagnetic in others, like Co-doped TiO$_2$~\cite{janisch06}.

The accepted effective theoretical
models~\cite{dietl00,dassarmaSSC03,timm03,macdonaldNatMat,jungwirthRMP06} for
FM in DMS can be roughly divided into two broad categories
depending on whether the carriers mediating the FM interaction between
the magnetic impurities are itinerant free carriers (i.e. holes in the
valence band or electrons in the conduction band) or localized (or
bound) carriers (in an impurity band, for example). For the itinerant free
carrier case, e.g. Ga$_{1-x}$Mn$_x$As in the optimally doped ($x
\approx 0.05$) situation with the nominally high $T_{\rm C}$ ($\sim
100-170$ K), the effective magnetic interaction producing FM
is thought to be of the Ruderman-Kittel-Kasuya-Yosida (RKKY)-type~\cite{dietl00,dassarmaSSC03,timm03,macdonaldNatMat,jungwirthRMP06,dassarma03}
leading to a mean-field FM transition by virtue of there being many
more effective impurity magnetic moments than free carriers. Such RKKY carrier-mediated mean-field FM seems to describe well~\cite{dietl00,macdonaldNatMat,jungwirthRMP06} the optimally doped (Ga,Mn)As regime. For the
localized case, where disorder should play an important role, the FM
transition is thought to be caused by the temperature-driven magnetic
percolation transition of bound magnetic polarons (BMP)~\cite{kaminski02},
objects where one localized carrier is magnetically strongly
correlated with a few neighboring magnetic impurities through the
local exchange interaction. An example of such BMP percolation induced FM is thought to be the localized impurity band Ge$_{1-x}$Mn$_x$ DMS system~\cite{li05}. Typically, the RKKY (BMP percolation)
ferromagnetic $T_{\rm C}$ is relatively high (low).

In this Letter, we argue that it is possible, perhaps even
likely in some impurity band systems, for these two mechanisms to operate together in
(at least some) DMS materials, leading to an intriguing situation
where the RKKY mechanism, mediated by {\em thermally excited free carriers},
dominates at high $T$, whereas the low $T$ regime (where thermal
carrier activation is exponentially suppressed) is dominated by the
polaron percolation mechanism in the impurity band.
By theoretically analyzing such an interesting multi-mechanism DMS situation using simple physical models, we show that it may be generically possible for DMS materials to have re-entrant FM with a paramagnetic state sandwiched between the high-temperature 'activated' RKKY-type and the low-temperature BMP percolation-type FM.

The system we consider is a localized impurity band system which is insulating at $T=0$. In this scenario, the density of band carriers is a function of temperature $T$ and  activation energy $\Delta$, 
\begin{equation}
n_c(T,\Delta)= n_{c0} \exp\left(-\Delta/k_B T \right)\,. 
\end{equation}
As a consequence, the RKKY mechanism is inefficient at low $T$ (typically for $T < \Delta$) when the free carrier density is too low, while the activated carriers can mediate FM at higher temperatures. Therefore, this 'activated' RKKY mechanism leads, by itself, to a FM phase between two disordered phases at high $T>T_1$ and low $T<T_2$ temperatures. At low $T$, other mechanisms not requiring free carriers (e.g. magnetic polaron percolation) can come into play producing a further magnetic transition with critical temperature $T_3$. When $T_3<T_2$ there are two distinct ferromagnetic phases and the system exhibits re-entrant FM. 

The Hamiltonian of the exchange interaction between magnetic impurities and the carriers is
$H=\sum_i Ja_0^3 \,\, {\bf S}_i \cdot {\bf s}({\bf R}_i)$
where $J$ is the local exchange coupling between the impurity spin ${\bf S}_i$ located at ${\bf R}_i$ and the carrier spin density ${\bf s} ({\bf r})$, and $a_0^3$ is the unit cell volume. 

\begin{figure}
	\centering
		\resizebox{55mm}{!}
		{\includegraphics{magn-nc03-ni3.eps}}
	\resizebox{57mm}{!}
		{\includegraphics{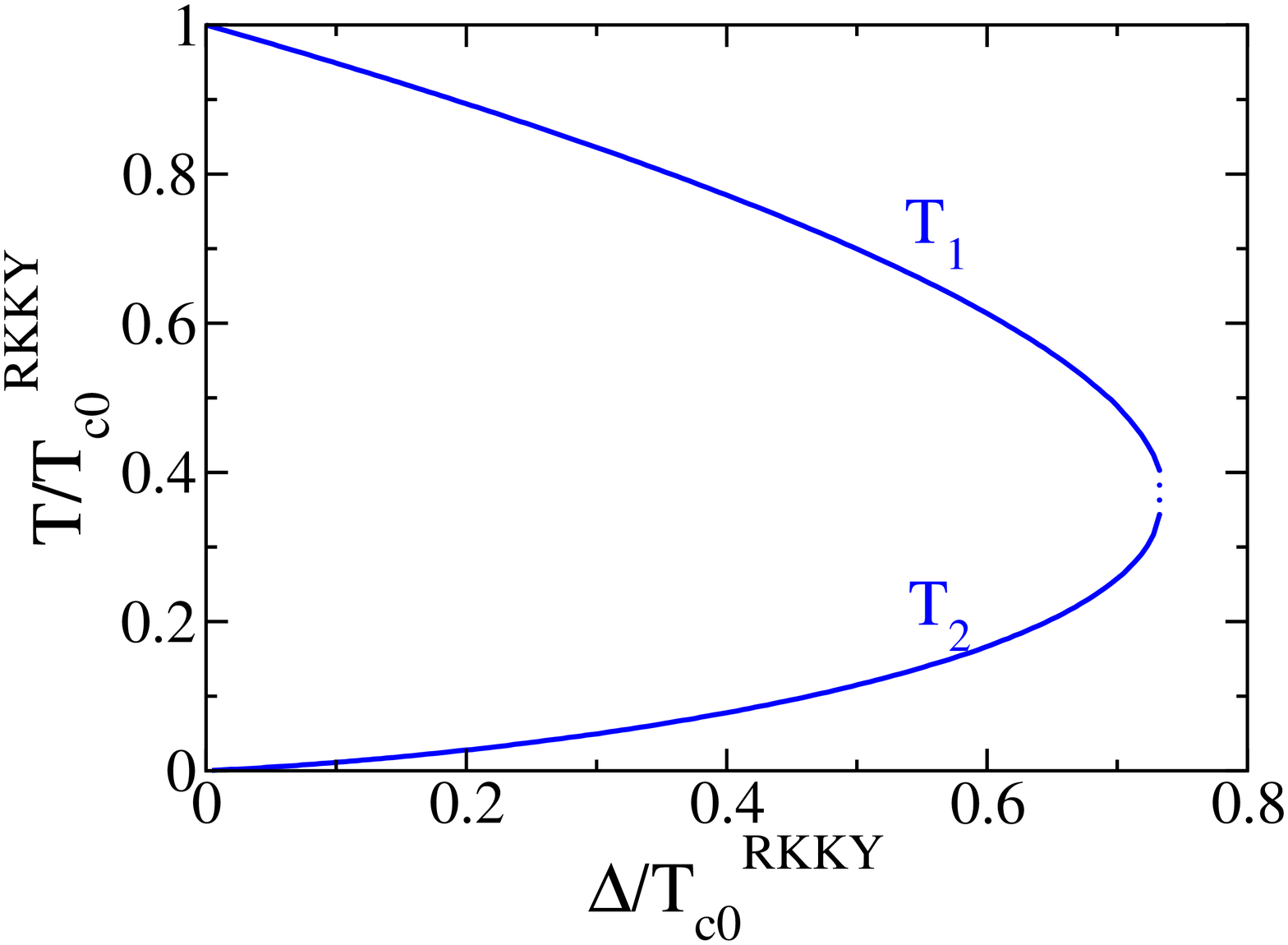}}
	\caption{(a) Impurity magnetization $\langle S_z \rangle/S$ corresponding to the 'activated' RKKY model, Eq.~(\ref{eq:RKKY}), for different values of the  carrier activation energy $\Delta$. The parameters used are $n_{c0}=3 \times 10^{20}$ cm$^{-3}$, $n_i= 3 \times 10^{21}$ cm$^{-3}$, $J=1$ eV.  (b) The two critical temperatures $T_1$ and $T_2$ of the 'activated' RKKY model, Eq.~(\ref{eq:RKKY}).  $T$ and $\Delta$ are in units of $T_{c0}^{RKKY}\, \equiv \,T_1 (\Delta=0)$. }
		\label{fig:RKKY}
\end{figure} 

As an illustration, we first discuss the hypothetical case of an intrinsic semiconductor with magnetic impurities which do not contribute carriers (therefore, there exists no impurity band where BMPs could form). In this case, the RKKY  mechanism can only be mediated by carriers thermally activated from the valence to the conduction band. Another mechanism that could play a role in this intrinsic case is the Bloembergen-Rowland mechanism~\cite{bloembergen55} mediated by virtual electron  excitations across the band gap. This mechanism was considered in the context of II-VI DMS~\cite{furdyna88} and was finally dismissed as being too weak compared to superexchange~\cite{larson88}. Therefore, we do not further include this mechanism in our discussion.

We now consider this 'activated' RKKY mechanism within a mean-field theory in the limit of nondegenerate carriers. In the activated carrier scenario we are studying, the Fermi energy is smaller than the temperature, and hence the limit of nondegenerate carriers is  appropriate. Within mean-field theory, the impurity spins act upon the carrier spins as an effective magnetic field $\propto Ja_0^3 n_i \langle S_z \rangle$ while the carrier spins act upon the impurity spins with an affective field  $\propto Ja_0^3 n_c \langle s_z \rangle$. As a result, the magnetization of the magnetic impurities $\langle S_z \rangle/S$ is calculated self-consistently giving 
\begin{equation}
\label{eq:RKKY}
{\frac {\langle S_z \rangle} {S}} = B_S \left[{\frac{J a_0^3 n_c(T,\Delta)}{k_B T}} s  B_s \left({\frac{J a_0^3 n_i \langle S_z \rangle}{k_B T}}\right)\right]\, ,
\end{equation}
where $B_s (y)$ is the standard  Brillouin function.  The magnetization is plotted in Fig.~\ref{fig:RKKY} (a) for three different values of $\Delta$. The $\Delta=0$ curve has a non-standard concave shape as expected for a low density of carriers~\cite{dassarma03}.  The critical temperatures are given by the points at which the magnetic susceptibility diverges. The susceptibility is essentially that of the magnetic impurities, as usually $n_c << n_i$ due to carrier compensation, and in the paramagnetic regions is given by $\chi (T) \equiv n_i \partial \left( g_i \mu_B \langle S_z \rangle \right)/\partial B $. 
For $\Delta=0$, there is only one critical temperature given by the standard result  $T_{c0}^{RKKY}={\frac{1}{3}} J a_0^3 \sqrt{n_{c0} n_i} \sqrt{(S+1) (s+1)}$. For $\Delta \ne 0$, the model gives two critical temperatures $T_1$ and $T_2$, shown in Fig.~\ref{fig:RKKY} (b): at low $T$, $n_c$ gets exponentially suppressed  and the localized moments are independent of each other; as $T$ increases, the band gets populated and the carrier-mediated FM kicks in at $T_2$, which increases with $\Delta$. At even higher $T$, thermal disorder produces the standard ferro-to-paramagnetic transition with critical temperature $T_1$, which decreases as $\Delta$ increases, due to the reduction of carrier density in the band. The $T_1$ and $T_2$ curves meet at $\Delta \approx 0.73\, T_{c0}^{RKKY}$ and, for $\Delta > 0.73\, T_{c0}^{RKKY}$, the density of carriers is always too low to mediate FM so the 'activated' RKKY model gives paramagnetism at all $T$. 
The curves for different parameters scale with $T_{c0}^{RKKY}\, \equiv \, T_1 (\Delta=0)$.

\begin{figure}
	\centering
	\resizebox{68mm}{!}
		{\includegraphics{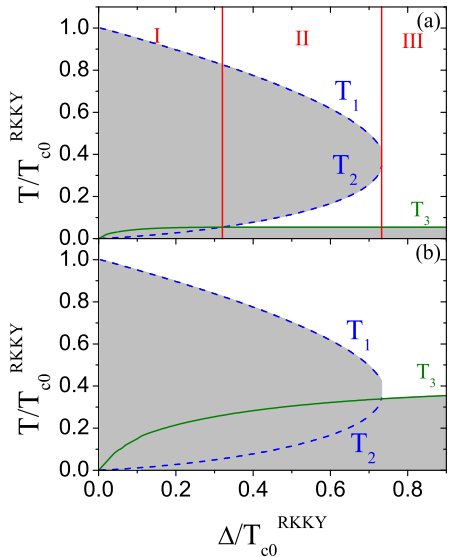}}
	\caption{Phase diagrams for the 'activated' RKKY-BMP percolation combined model in two typical cases. The dashed lines represent the two critical temperatures of the 'activated' RKKY model, Eq.~(\ref{eq:RKKY}). The solid line is the critical temperature in the bound magnetic polaron percolation model given by Eq.~(\ref{eq:perc}). The shaded areas are the ferromagnetic phases and the white ones are paramagnetic.
The parameters used are $n_i= 3 \times 10^{21}$ cm$^{-3}$, $J=1$ eV, $a_0=3.23 {\rm \AA}$ and (a) $n_{c0}=3 \times 10^{19}$ cm$^{-3}$,  $a_B=4 {\rm \AA}$  ($ a_B^3 n_{c0}\approx 0.002$),  (b) $n_{c0}=1 \times 10^{20}$ cm$^{-3}$,  $a_B=5 {\rm \AA}$ ($ a_B^3 n_{c0} \approx 0.0125$).  }
\label{fig:RKKY-perc}
\end{figure}

We consider now the more realistic case in which there is an impurity band. The carriers in the impurity band can come from magnetic impurities acting as dopants, like in the III-V semiconductors, or from other dopants, like the oxygen vacancies which act as shallow donors in the magnetic oxides.  We assume that, at $T=0$ the conduction (or valence) band is empty so all the free carriers in the band are produced by thermal activation from the impurity band. At low $T$, the carriers in the impurity band can be strongly localized and mediate FM through the formation of BMPs which grow as $T$ lowers, finally overlapping in clusters and producing a FM transition at percolation~\cite{kaminski02}. In our model of thermally activated carriers, the density of carriers in the impurity band is  $n_c^*(T,\Delta)=n_{c0}-n_c= n_{c0}\left[1-\exp(-\Delta/k_B T)\right]$, where $n_c$ is the density of  carriers in the valence or conduction band. $\Delta$ is the smallest energy gap for electron excitations, which is given by the activation energy of a carrier at the impurity level.  
The critical temperature in the BMP percolation model is given by~\cite{kaminski02}
\begin{equation}
\label{eq:perc}
T_c^{\rm perc} = s S J \left({\frac{a_0}{a_B}}\right)^3 (a_B^3 n_c^*)^{1/3} \sqrt{\frac{n_i}{n_c^*}} \,\, e^{-\frac{0.86}{(a_B^3 n_c^*)^{1/3}}} \, ,
\end{equation}
where $a_B$ is the carrier localization radius, and $n_c^*$ is the carrier density at $T_c^{\rm perc}$. Eq.~(\ref{eq:perc}) is valid in the low carrier density limit $ a_B^3 n_c^*<< 1$. As the density of carriers involved in polaron formation $n_c^*$ increases with $\Delta$, $T_c^{\rm perc}$ also increases, saturating for large values of $\Delta$, when $n_c^* \approx n_{c0}$, as shown in Fig.~\ref{fig:RKKY-perc} (solid line). The value of $T_c^{\rm perc}$ is mainly dominated by the value of $a_B^3 n_c^*$, on which it depends exponentially. Another mechanism that could arise when there are no free carriers is the Bloembergen-Rowland-type mechanism proposed in Ref.~\cite{litvinov01} in the context of Mn-doped III-V semiconductors, where the virtual electron excitations take place between the impurity levels and the band. We do not expect, however, this mechanism to be stronger than the BMP percolation, and therefore, it does not affect our conclusions.

\begin{figure}
	\centering
	\resizebox{65mm}{!}
		{\includegraphics{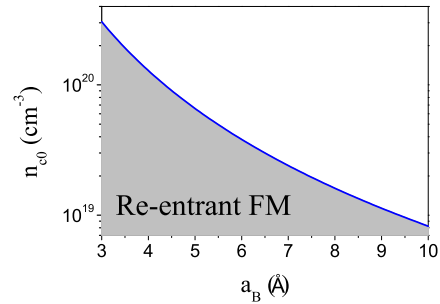}}
	\caption{The shaded area gives the values of $a_B$ and $n_{c0}$ compatible with the re-entrant FM behavior illustrated in Fig.~\ref{fig:RKKY-perc}(a). The curve is given by $a_B^3 n_{c0} = 0.00824$.  }
\label{fig:aB-nc0}
\end{figure}

In Fig.~\ref{fig:RKKY-perc} we show two typical phase diagrams resulting from the interplay of the 'activated' RKKY and the BMP percolation models. The parameters chosen here are representative of real systems. We have fixed the magnetic impurity concentration to $x=0.1$, $J=1$ eV, and $S=1$. The other two main parameters left are $n_{c0}$ and $a_B$. $n_{c0}$ is chosen so that $n_{c0}/n_i \leq 0.1$, as it is usually found in DMS systems due to strong carrier compensation. Finally, $a_B$ is chosen to fulfill 
the applicability conditions of Eq.~(\ref{eq:perc}). 
There are two qualitatively different phase diagrams that can occur depending on the value of $a_B^3 n_{c0}$. For the lowest values of $a_B^3 n_{c0}$, $T_3 \,\equiv\,T_c^{\rm perc}$ is relatively small, and we get the novel scenario depicted in Fig.~\ref{fig:RKKY-perc} (a): in region I, corresponding to $\Delta/T_{c0}^{\rm RKKY} \lesssim 0.32$ for the parameter values chosen in this figure, the system is FM with critical temperature $T_1$; in region III, with $\Delta/T_{c0}^{\rm RKKY} \gtrsim 0.73$ for any set of parameters, the system is ferromagnetic with $T_c=T_3$; finally, in region II, which corresponds to intermediate values of $\Delta$ ($0.32 \lesssim \Delta/T_{c0}^{\rm RKKY}  \lesssim 0.73$), $T_3 <T_2$ and the system shows re-entrant FM with three distinct critical temperatures $T_1$, $T_2$, and $T_3$. For larger values of $a_B^3 n_{c0}$ (but still $<< 1$), as in Fig.~\ref{fig:RKKY-perc} (b), $T_3>T_2$ for all values of $\Delta$ and the system is ferromagnetic for  $T < \max(T_1, T_3)$.   

Although our theoretical analysis, being physically motivated, is formally correct, the question naturally arises about the observability of our predicted re-entrant FM behavior in DMS materials. There are three general conditions that should be met: (i) at $T=0$ there should be no carriers in the conduction (or valence) band so all the carriers mediating RKKY come from thermal activation; (ii) for the scenario illustrated in Fig.~\ref{fig:RKKY-perc} (a) to occur we need to fulfill the condition $T_c^{\rm perc}(n_{c0}) < T_2^{\rm max} \approx 0.3 \times T_{c}^{\rm RKKY} (n_{c0})$; and (iii) we also have to ensure that the system is in region II in Fig.~\ref{fig:RKKY-perc} (a) to actually observe re-entrant FM.

Condition (i) basically implies that the system has to have activated-like resistivity 
(decreasing as T increases) limiting the suitable systems to those with so-called 'insulating behavior', for example, very lightly doped (Ga,Mn)As, (In,Mn)As, and magnetic oxides. Condition (ii) puts restrictions on the relative values of $a_B$ and $n_{c0}$. The equation in (ii) is independent of $J$ and $n_i$ and reduces to the inequality $a_B^3 n_{c0} < 0.00824$ for $S=1$. This result is illustrated in Fig.~\ref{fig:aB-nc0} where the curve is given by the maximum allowed value of the product $a_B^3 n_{c0}$. Below this curve we get the scenario  in Fig.~\ref{fig:RKKY-perc} (a) while the scenario in Fig.~\ref{fig:RKKY-perc} (b) occurs for $a_B$ and $n_{c0}$ values above the curve.  
The carrier confinement radius is given by $a_B=\epsilon (m/m^*)a$ where $a=0.52$ \AA, $\epsilon$ is the semiconductor dielectric constant, and $m^*$ is the polaron mass (usually $m^*/m > 1$).  For diluted magnetic semiconductors and magnetic oxides we generally expect $3$\AA$< a_B < 10$\AA. Hence, as shown in Fig.~\ref{fig:aB-nc0}, which is the important materials phase diagram to keep in mind in searching for a suitable system to observe our predicted re-entrant FM, we need to keep a relatively low density of carriers, but well within experimentally achievable values. 

Condition (iii) can be fulfilled, in general terms, when $\Delta$ is comparable to $T_c$. This rules out Ga$_{1-x}$Mn$_x$As, with $\Delta/T_c \sim 10$ ($T_c \sim 100$ K, and $\Delta \simeq 110$ meV~\cite{schairer74}), which would rather be in region III (for very low carrier density samples). On the other hand, the re-entrant FM scenario might be applicable in In$_{1-x}$Mn$_x$As, which has very low activation energies (see, for example, Fig. 2 in Ref.~\cite{dietlreview02}) and low $T_c \sim 50$ K. However, the most likely candidate for observing our predicted re-entrant FM is possibly a doped magnetic oxide material, e.g. Ti$_{1-x}$Co$_x$O$_2$ (with $30 {\rm meV} < \Delta < 70$ meV~\cite{shinde} and very similar critical temperatures $T_c \sim 700$ K), which is an insulator at low $T$, but exhibits essentially $T$-independent Drude transport behavior, due to thermally activated carriers, at room temperature or above. In such a system, it is quite possible that the high-$T$  FM is mediated by thermally activated carriers, whereas at low $T$ a BMP percolation FM takes over~\cite{coey05} as in region I of Fig.~\ref{fig:RKKY-perc} (a). 

We have neglected the direct exchange interaction between the magnetic impurities, which is short-ranged, and could also play a role, particularly when the density of impurities $n_i$ is large, completely destroying the lower FM phase~\cite{kaminski04}, and leading to a spin-glass phase. These spin-glass low-$T$ phases have been observed in II-VI DMS~\cite{furdyna88}, and (Ga,Mn)N~\cite{dhar03}, in samples that never show FM order. More relevant to our case, a suppresion of magnetization at low $T$ in zero field cooled curves has been reported 
in the magnetic oxide V-doped ZnO and possibly in other magnetic oxides~\cite{hong05}. In principle, therefore, direct antiferromagnetic exchange between the magnetic dopant impurities could compete with (or even suppress) our predicted low-T re-entrant FM phase, but the origin of this direct exchange being completely different from the carrier-mediated mechanisms producing the re-entrant FM behavior itself, we find it difficult to believe that such a suppression of re-entrance can be generic. We suggest detailed $T$-dependent magnetization studies in the shaded region of our Fig.~\ref{fig:aB-nc0} to search for our predicted re-entrant FM behavior.

We have considered, using physically motivated effective models of FM in DMS materials, the intriguing possibility of generic re-entrant FM in an insulating class of DMS systems. The models we use are considered to be successful minimal carrier-mediated models~\cite{dietl00,dassarmaSSC03,timm03,macdonaldNatMat,jungwirthRMP06} for understanding FM behavior and predicting ferromagnetic $T_{\rm C}$ in itinerant and localized DMS materials. The new idea in this work has been to point out that these 'competing' FM mechanisms, e.g. 'activated' RKKY and BMP percolation, could, in principle, exist together in a single sample where thermal activation leads to a high-$T$ effective free-carrier mechanism mediating the RKKY ferromagnetism, and the low-$T$ FM, where thermal activation of free carriers is exponentially suppressed, is mediated by localized bound carriers in an impurity band through the polaron percolation mechanism. We show that, depending on the materials parameters, such a situation with competing high-$T$ and low-$T$ FM mechanisms generically allows for re-entrant FM with an intermediate-temperature anomalous paramagnetic phase intervening between the higher temperature RKKY FM phase and the lower temperature BMP percolation FM phase. More experimental work in DMS materials is needed to confirm the existence of re-entrant FM, but the phenomenon should exist on firm theoretical grounds.

This work is supported by the NSF and the NRI SWAN program.

\bibliography{reentrance}

\begin{thebibliography}{23}
\expandafter\ifx\csname natexlab\endcsname\relax\def\natexlab#1{#1}\fi
\expandafter\ifx\csname bibnamefont\endcsname\relax
  \def\bibnamefont#1{#1}\fi
\expandafter\ifx\csname bibfnamefont\endcsname\relax
  \def\bibfnamefont#1{#1}\fi
\expandafter\ifx\csname citenamefont\endcsname\relax
  \def\citenamefont#1{#1}\fi
\expandafter\ifx\csname url\endcsname\relax
  \def\url#1{\texttt{#1}}\fi
\expandafter\ifx\csname urlprefix\endcsname\relax\def\urlprefix{URL }\fi
\providecommand{\bibinfo}[2]{#2}
\providecommand{\eprint}[2][]{\url{#2}}

\bibitem[{\citenamefont{Ohno et~al.}(1996)\citenamefont{Ohno, Shen, Matsukura,
  Oiwa, Endo, Katsumoto, and Iye}}]{ohno96}
\bibinfo{author}{\bibfnamefont{H.}~\bibnamefont{Ohno et al}},
  \bibinfo{journal}{Appl. Phys. Lett.} \textbf{\bibinfo{volume}{69}},
  \bibinfo{pages}{363} (\bibinfo{year}{1996}).

\bibitem[{\citenamefont{Matsumoto et~al.}(2001)\citenamefont{Matsumoto,
  Murakami, Shono, Hasegawa, Fukumura, Kawasaki, Ahmet, Chikyow, Koshihara, and
  Koinuma}}]{matsumoto01}
\bibinfo{author}{\bibfnamefont{Y.}~\bibnamefont{Matsumoto et al}},
  \bibinfo{journal}{Science} \textbf{\bibinfo{volume}{291}},
  \bibinfo{pages}{854} (\bibinfo{year}{2001}).

\bibitem[{\citenamefont{Furdyna}(1988)}]{furdyna88}
\bibinfo{author}{\bibfnamefont{J.}~\bibnamefont{Furdyna}}, \bibinfo{journal}{J.
  Appl. Phys.} \textbf{\bibinfo{volume}{64}}, \bibinfo{pages}{R29}
  (\bibinfo{year}{1988}).

\bibitem[{\citenamefont{von Molnar et~al.}(1991)\citenamefont{von Molnar,
  Munekata, Ohno, and Chang}}]{vonmolnar91}
\bibinfo{author}{\bibfnamefont{S.}~\bibnamefont{von Molnar et al}},
  \bibinfo{journal}{J. Magn. Magn. Mater.} \textbf{\bibinfo{volume}{93}},
  \bibinfo{pages}{356} (\bibinfo{year}{1991}).

\bibitem[{\citenamefont{Larson et~al.}(1988)\citenamefont{Larson, Hass,
  Ehrenreich, and Carlsson}}]{larson88}
\bibinfo{author}{\bibfnamefont{B.~E.} \bibnamefont{Larson et al}},
\bibinfo{journal}{Phys. Rev. B}
  \textbf{\bibinfo{volume}{37}}, \bibinfo{pages}{4137} (\bibinfo{year}{1988}).

\bibitem[{\citenamefont{Janisch and Spaldin}(2006)}]{janisch06}
\bibinfo{author}{\bibfnamefont{R.}~\bibnamefont{Janisch}} \bibnamefont{and}
  \bibinfo{author}{\bibfnamefont{N.~A.} \bibnamefont{Spaldin}},
  \bibinfo{journal}{Phys. Rev. B} \textbf{\bibinfo{volume}{73}},
  \bibinfo{pages}{035201} (\bibinfo{year}{2006}).

\bibitem[{\citenamefont{Dietl et~al.}(2000)\citenamefont{Dietl, Ohno,
  Matsukura, Cibert, and Ferrand}}]{dietl00}
\bibinfo{author}{\bibfnamefont{T.}~\bibnamefont{Dietl et al}},
  \bibinfo{journal}{Science} \textbf{\bibinfo{volume}{287}},
  \bibinfo{pages}{1019} (\bibinfo{year}{2000}).

\bibitem[{\citenamefont{{Das Sarma}
  et~al.}(2003{\natexlab{a}})\citenamefont{{Das Sarma}, Hwang, and
  Kaminski}}]{dassarmaSSC03}
\bibinfo{author}{\bibfnamefont{S.}~\bibnamefont{{Das Sarma} et al}},
  \bibinfo{journal}{Solid State Communications} \textbf{\bibinfo{volume}{127}},
  \bibinfo{pages}{99} (\bibinfo{year}{2003}{\natexlab{a}}).

\bibitem[{\citenamefont{Timm}(2003)}]{timm03}
\bibinfo{author}{\bibfnamefont{C.}~\bibnamefont{Timm}}, \bibinfo{journal}{J.
  Phys.: Condens. Matter} \textbf{\bibinfo{volume}{15}}, \bibinfo{pages}{R1865}
  (\bibinfo{year}{2003}).

\bibitem[{\citenamefont{MacDonald et~al.}(2005)\citenamefont{MacDonald,
  Schiffer, and Samarth}}]{macdonaldNatMat}
\bibinfo{author}{\bibfnamefont{A.}~\bibnamefont{MacDonald et al}},
  \bibinfo{journal}{Nature Materials} \textbf{\bibinfo{volume}{4}},
  \bibinfo{pages}{195} (\bibinfo{year}{2005}).

\bibitem[{\citenamefont{Jungwirth et~al.}(2006)\citenamefont{Jungwirth, Sinova,
  Masek, Kucera, and MacDonald}}]{jungwirthRMP06}
\bibinfo{author}{\bibfnamefont{T.}~\bibnamefont{Jungwirth et al}},
  \bibinfo{journal}{Rev. Mod. Phys.} \textbf{\bibinfo{volume}{78}},
  \bibinfo{pages}{809} (\bibinfo{year}{2006}).

\bibitem[{\citenamefont{{Das Sarma}
  et~al.}(2003{\natexlab{b}})\citenamefont{{Das Sarma}, Hwang, and
  Kaminski}}]{dassarma03}
\bibinfo{author}{\bibfnamefont{S.}~\bibnamefont{{Das Sarma} et al}},
  \bibinfo{journal}{Phys. Rev. B} \textbf{\bibinfo{volume}{67}},
  \bibinfo{pages}{155201} (\bibinfo{year}{2003}{\natexlab{b}}).

\bibitem[{\citenamefont{Kaminski and {Das Sarma}}(2002)}]{kaminski02}
\bibinfo{author}{\bibfnamefont{A.}~\bibnamefont{Kaminski}} \bibnamefont{and}
  \bibinfo{author}{\bibfnamefont{S.}~\bibnamefont{{Das Sarma}}},
  \bibinfo{journal}{Phys. Rev. Lett.} \textbf{\bibinfo{volume}{88}},
  \bibinfo{pages}{247202} (\bibinfo{year}{2002}).

\bibitem[{\citenamefont{Li et~al.}(2005)\citenamefont{Li, Wendelken, Shen,
  Feldman, Thompson, and Weitering}}]{li05}
\bibinfo{author}{\bibfnamefont{A.~P.} \bibnamefont{Li et al}},
\bibinfo{journal}{Phys. Rev. B}
  \textbf{\bibinfo{volume}{72}}, \bibinfo{pages}{195205}
  (\bibinfo{year}{2005}).

\bibitem[{\citenamefont{Bloembergen and Rowland}(1955)}]{bloembergen55}
\bibinfo{author}{\bibfnamefont{N.}~\bibnamefont{Bloembergen}} \bibnamefont{and}
  \bibinfo{author}{\bibfnamefont{T.~J.} \bibnamefont{Rowland}},
  \bibinfo{journal}{Phys. Rev.} \textbf{\bibinfo{volume}{97}},
  \bibinfo{pages}{1679} (\bibinfo{year}{1955}).

\bibitem[{\citenamefont{Litvinov and Dugaev}(2001)}]{litvinov01}
\bibinfo{author}{\bibfnamefont{V.~I.} \bibnamefont{Litvinov}} \bibnamefont{and}
  \bibinfo{author}{\bibfnamefont{V.~K.} \bibnamefont{Dugaev}},
  \bibinfo{journal}{Phys. Rev. Lett.} \textbf{\bibinfo{volume}{86}},
  \bibinfo{pages}{5593} (\bibinfo{year}{2001}).

\bibitem[{\citenamefont{Schairer and Schmidt}(1974)}]{schairer74}
\bibinfo{author}{\bibfnamefont{W.}~\bibnamefont{Schairer}} \bibnamefont{and}
  \bibinfo{author}{\bibfnamefont{M.}~\bibnamefont{Schmidt}},
  \bibinfo{journal}{Phys. Rev. B} \textbf{\bibinfo{volume}{10}},
  \bibinfo{pages}{2501} (\bibinfo{year}{1974}).

\bibitem[{\citenamefont{Dietl}(2002)}]{dietlreview02}
\bibinfo{author}{\bibfnamefont{T.}~\bibnamefont{Dietl}},
  \bibinfo{journal}{Semicond. Sci. Technol.} \textbf{\bibinfo{volume}{17}},
  \bibinfo{pages}{377} (\bibinfo{year}{2002}).

\bibitem[{shi()}]{shinde}
\bibinfo{note}{S.R. Shinde {\it et al} (unpublished)}.

\bibitem[{\citenamefont{Coey et~al.}(2005)\citenamefont{Coey, Venkatesan, and
  Fitzgerald}}]{coey05}
\bibinfo{author}{\bibfnamefont{J.}~\bibnamefont{Coey et al}},
  \bibinfo{journal}{Nature Materials} \textbf{\bibinfo{volume}{4}},
  \bibinfo{pages}{173} (\bibinfo{year}{2005}).

\bibitem[{\citenamefont{Kaminski et~al.}(2004)\citenamefont{Kaminski, Galitski,
  and {Das Sarma}}}]{kaminski04}
\bibinfo{author}{\bibfnamefont{A.}~\bibnamefont{Kaminski et al}},
, \bibinfo{journal}{Phys. Rev. B} \textbf{\bibinfo{volume}{70}},
  \bibinfo{pages}{115216} (\bibinfo{year}{2004}).

\bibitem[{\citenamefont{Dhar et~al.}(2003)\citenamefont{Dhar, Brandt, Trampert,
  Friedland, Sun, and Ploog}}]{dhar03}
\bibinfo{author}{\bibfnamefont{S.}~\bibnamefont{Dhar et al}},
  \bibinfo{journal}{Phys. Rev. B} \textbf{\bibinfo{volume}{67}},
  \bibinfo{pages}{165205} (\bibinfo{year}{2003}).

\bibitem[{\citenamefont{Hong et~al.}(2005)\citenamefont{Hong, Sakai, and
  Hassini}}]{hong05}
\bibinfo{author}{\bibfnamefont{N.}~\bibnamefont{Hong et al}},
  \bibinfo{journal}{J. Phys.: Condens. Matter} \textbf{\bibinfo{volume}{17}},
  \bibinfo{pages}{199} (\bibinfo{year}{2005}).

\end{thebibliography}

\end{document}